\documentclass[final,3p,times,twocolumn]{elsarticle}

\usepackage{amssymb}
\usepackage{epsfig}
\usepackage{color}
\usepackage{slashed}
\usepackage{hyperref}

\newcommand{\ar}{\arrowvert} 
\newcommand{\ra}{\rangle} 
\newcommand{\la}{\langle}

\newcommand{\be}{\begin{equation}} 
\newcommand{\ee}{\end{equation}} 
\newcommand{\ba}{\begin{eqnarray}} 
\newcommand{\ea}{\end{eqnarray}}
\newcommand{\x}{x}

\journal{Nuclear Physics A}
\begin{document}
\begin{frontmatter}
\title{Soft interactions in jet quenching}

\author[chd]{Carlos Hidalgo-Duque}
\author[fjle]{Felipe J. Llanes-Estrada}
\address[chd]{Instituto de F\'isica Corpuscular (IFIC),
             CSIC-Universidad de Valencia,
             Institutos de Investigaci\'on de Paterna,
             Aptd. 22085, E-46071 Valencia, Spain.}
\address[fjle]{ Departamento de F\'isica Te\'orica I, Universidad Complutense, 28040 Madrid, Spain.}

\date{\today}

\begin{abstract}
\rule{0ex}{3ex}
We study the collisional aspects of jet quenching in a high energy nuclear collision, especially in the final state pion gas. The jet has a large energy, and acquires momentum transverse to its axis more effectively by multiple soft collisions than by few hard scatterings (as known from analogous systems such as $J/\psi$ production at Hera). Such regime of large E and small momentum transfer 
corresponds to Regge kinematics and is
characteristically dominated by the pomeron. From this insight we estimate the jet quenching parameter in the hadron medium (largely a pion gas) at the end of the collision, which is naturally small and increases with temperature in line with the gas density. The physics in the quark-gluon plasma/liquid phase is less obvious, 
and here we revisit a couple of simple estimates that suggest indeed that the pomeron-mediated interactions are very relevant 
and should be included in analysis of
the jet quenching parameter.
Finally, the ocasional hard collisions produce features characteristic of a L\`{e}vy flight in the ${\bf q}_\perp^2$ plane perpendicular to the jet axis. We suggest one- and two-particle ${\bf q}_\perp$ correlations as interesting experimental probes.
\end{abstract}
\end{frontmatter}

\section{Introduction}

Bjorken is usually credited~\cite{Bjorken82} with having pointed out for the first time that jets would lose significant energy while plowing through a hot hadron medium. His treatment of the jet parton collisions with the medium quarks and gluons was perturbative, and guessed IR and UV cutoffs were imposed. 

Bjorken's work was improved later on by properly obtaining the cutoffs
with the natural in-medio regularization, the phenomenology further explored, and the study extended to include heavy quarks~\cite{Gyulassy:1990ye,Thoma:1990fm,Braaten:1991we}.

Among the many later works, several well-known ones highlighted that braking radiation (perhaps off-shell) following the collisions of the jets with the medium partons would be more important in shedding energy than the actual collisions, and a treatment analogous to the Landau-Pomeranchuk-Migdal theory in QED was provided~\cite{Baier:1996sk}

Nowadays, the generally accepted picture is  a sequence of parton--medium collisions, in which the jet acquires transverse momentum ${\bf q}_\perp$, accompanied by emission of braking radiation, that enhances the energy loss per unit length $dE/dx$. The emission of collinear partons in the jet and the running of the in-medio fragmentation functions seem to be under good theoretical control as they depend on a hard scale, the parton virtuality $Q^2$~\cite{Renk,Domdey,Armesto} and are thus amenable to perturbative treatment.

Often, the collisional part is described by a single jet-quenching parameter
 $\hat{q}$ that represents the increase in squared transverse momentum of a particle in the jet per unit length in the medium,
\begin{equation} \label{defq}
\hat{q} = \frac{\Delta q_{\perp}^{2}}{\Delta \lambda}\ .
\end{equation}
The quick kinetic-theory estimate broadly used is based on the mean-free path
$\lambda = \frac{1}{n \sigma}$, with $n$ and $\sigma$ the medium density of targets and the total parton-target cross section, respectively,
\begin{equation} \label{defqhat}
\hat{q} = \left<\Delta q_{\perp}^{2}\right> ~ n~\sigma \ .
\end{equation}

The experimental observation of jet quenching in central collisions seems to be confirmed now by the LHC experiments~\cite{CMSdijets}, even to high $P_t$ of about 350 GeV.
Earlier PHENIX data suggests that the jet quenching parameter $\hat{q}$ has to be of order 10-15 GeV$^2/$fm~\cite{Adare:2008cg}. 
Values around 14-15 GeV$^2/$fm were also found in~\cite{Dainese:2004te}.

These values were larger than typical estimates (that are mostly perturbative, two of the simplest are redrawn below).
Most-recent extractions, such as those compiled by the JET collaboration~\cite{Burke:2013yra} are significantly smaller, with $\hat{q}\propto T^3$ and $\hat{q}~ (350{\rm MeV}-450 {\rm MeV}) = 1-3$GeV$^2/$fm.

 Many authors are now aware that a non-perturbative extraction is necessary, but it is not totally clear how to proceed to understand the data:  a  roadmap to $\hat{q}$ from theory is not agreed upon. We wish to make a modest contribution to the discussion by pointing out that Regge phenomenology is perhaps a good starting point, and estimate the contribution of elastic, diffractive and total cross sections in the final state of a pion gas, obtaining what should be a fair idea of the values that one can expect for $\hat{q}$ in that phase~\footnote{We find that values currently used by other researchers, of order $\hat{q}\sim 0.02 $GeV$^2/$fm in the pion gas, are a gross underestimate.
}.
This we compare to some simple existing computations in the quark-gluon plasma, with no claim for completeness there.

We can draw an analogy with well-known physics. Consider the inelastic photoproduction of the $J/\psi$ at Hera analyzed for example by the H1~\cite{Aaron:2010gz} collaboration. The reaction 
$\gamma p\to J/\psi X$ features an energetic photon converting to a charmonium (this would be the analogous of the incoming and outgoing jets after a collision with a fluid element) and a proton breaking up (analogous to the hadron medium).

\begin{figure}
\begin{center}
\includegraphics[width=0.50\textwidth]{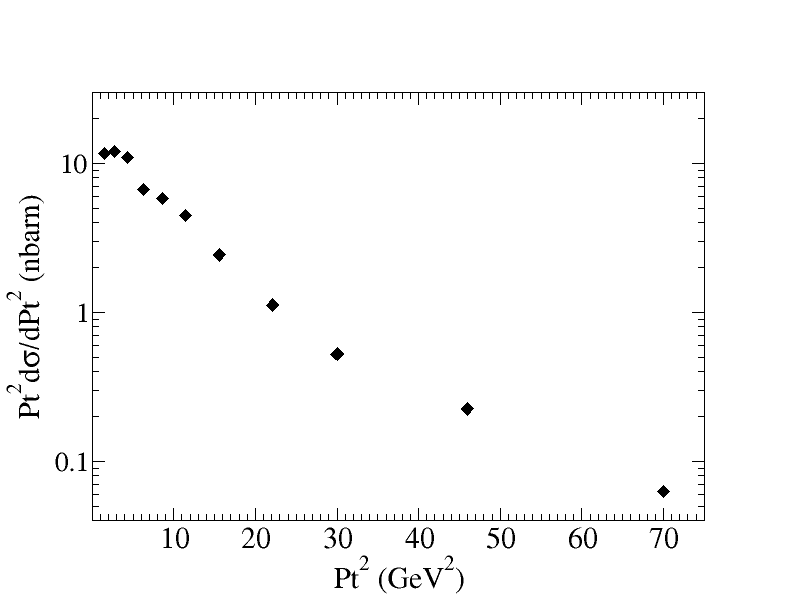}
\end{center}
\caption{Our rendering of Hera data~\cite{Aaron:2010gz} for the $P_t^2$-dependence of $J/\psi$ production in $\gamma~p \rightarrow J/\psi ~X$. 
We plot the central values of the differential cross-section weighted by the central values of $P_t^2$. This is a toy representation of the $P_t^2$- distribution acquired by the $J/\psi$, and is clearly peaked at low-$P_t$.}
\label{Hera_Data_1}
\end{figure}

In figure~\ref{Hera_Data_1} we replot the H1 data. 
We ask ourselves what is the average $P_t^2$ given to the $J/\psi$, in analogy with the collisional part of the jet quenching phenomenon. The figure shows that the $P_t$ distribution peaks below 1 GeV where the data stops. Thus, soft Regge physics is a better starting point than pQCD to understand the $P_t^2$ transfer.

\begin{figure}
\begin{center}
\includegraphics[width=0.30\textwidth]{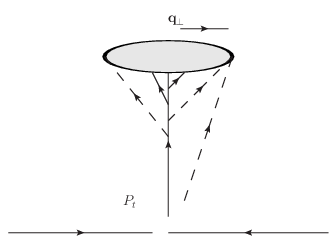}
\end{center}
\caption{We reserve $P_t$ for the total momentum of the jet measured perpendicularly to the {\emph{collision axis}}, while ${\bf q}_\perp$ denotes the momentum of each of the jet particles perpendicularly to the {\emph{jet axis}}. While the jet is composed of only the hard initial parton, a $\Delta {\bf q}_\perp$ kick implies a $P_t$ change (later redefining ${\bf q}_\perp$ to zero respect to the new jet axis), but for more particles, several ${\bf q}_\perp$ acquisitions may imply no variation of $P_t$ but rather a spread of the jet cone.}
\label{fig:variables}
\end{figure}

In the rest of the article we will address jet quenching from the point of view of high-energy, low-momentum transfer reactions (dominated by pomeron exchange), and compare with perturbation theory for a benchmark. Sections~\ref{sec:hadrons} and~\ref{sec:hadrons2} briefly discuss the hadron gas formed in the later stage of the collision.
 Section~\ref{sec:plasma} presents some very rough considerations about the quark-gluon plasma. 
We do not put much stress there, as the literature is vast, but rather employ a couple of examples to put our computation within the pion gas into context. 
In section~\ref{sec:correlations} we propose to employ 1- and 2-particle ${\bf q}_\perp^2$ observables to experimentally learn about mechanisms for acquisition of transverse momentum, whether hard or soft, and observe that the emergence of the ${\bf q}_\perp^2$ distribution can be simulated as a random walk with features characteristic of a L\'evy flight (multiple soft collisions with an ocasional  hard collision). We reserve $P_t$ for the momentum of the jet transverse to the collision axis, and ${\bf q}_\perp$ to the momentum of each particle in the jet perpendicular to its axis as depicted in figure~\ref{fig:variables} (of course, at early stages while the jet is composed of only one parton, they are equivalent).

We wrap the discussion up in section~\ref{sec:conc} where several additional example numerical computations are mentioned.

\section{Passage of a fast pion through the hadron medium~\label{sec:hadrons}}
The final stage of a heavy ion nuclear collision is a hadron gas~\cite{Prakash:1993bt}. The ratio of pion to either kaon or nucleon counts is normally 10:1, so  one can roughly talk of a pion gas. Following updated transport coefficients in the gas (viscosities, conductivities, etc.) by means of low-energy methods, such as chiral perturbation theory and its unitarization, and heavy-quark effective theory~\cite{Abreu:2012et,Torres-Rincon:2012sda,Abreu:2011ic,Dobado:2011qu,Dobado:2003wr,FernandezFraile:2005ka,FernandezFraile:2008vu}, 
we now undertake an assesment of this high-energy transport coefficient, the jet-quenching parameter in the hadron medium. The physics involved is partly different, as the jet is a high-energy probe, whereas all the various diffusion coefficients, viscosities and conductivities, involved slow relaxation of separations from equilibrium at the hadron scale; the cross-section therefore comes from different principles, and the correct approach is indeed to adopt Regge theory. 

 We take the pion number density to be in equilibrium with temperature $T=1/\beta$ and pion chemical potential $\mu < M_{\pi}$ (that can be introduced at low temperatures because the low-energy pion-pion interactions are practically elastic through $E=1.2$ GeV),
\ba \label{BoseEinstein}
n(T) = g \int \frac{d^{3}\vec{k}}{(2\pi)^{3}} \frac{1}{e^{\beta(E_{k} - \mu)}-1} 
\ .
\ea
It characteristically grows with the temperature as $T^3$.
For the pion ($\pi^-,\pi^0,\pi^+$) gas, the degeneracy factor is $g=3$. 

First we consider the interaction of one single fast pion entering the gas.
Its ${\bf q}_\perp$ transfer perfectly fits the  traditional definition of $\hat{q}$, referred to precisely one particle for which we will use $\hat{q}^{(1)}$.
It will also prove useful to introduce a separate variable $\hat{q}^{(N)}\simeq N \hat{q}^{(1)}$ with exactly the same meaning, but referred to $N$ particles~\footnote{The opening of the jet cone is of order $\hat{q}^{(N)}/P_t$ for low $\hat{q}^{(N)}$ and rather of order $\sqrt{\hat{q}^{(N)}} / P_t$ for large $\hat{q}^{(N)}$,  since each kicked particle is pushed in a random transverse direction often cancelling other side motions.}.
In the early stages of jet formation, when only one parton carries the entire momentum of the jet, $\hat{q}^{(N)}=\hat{q}^{(1)}=\hat{q}$, but later on $\hat{q}^{(N)}>\hat{q}^{(1)}$. 

In each collision, the elastic (and also the diffractive) contribution to transverse momentum transfer $\Delta {\bf q}_{\perp}^{2}$ is given, at first order in 
${\bf q}_\perp^2$, by
\be
t \simeq - \Delta {\bf q}_{\perp}^{2}\ .
\ee
We can average this easily if the cross section is known,
\begin{equation}
 \left<\Delta {\bf q}_{\perp}^{2}\right> = \frac {\int dt \frac{d\sigma} {dt} \Delta {\bf q}_{\perp}^{2} } {\int dt \frac{d\sigma} {dt} }\ .
\end{equation}
This is the quantity to be substituted in Eq.~(\ref{defqhat}) for the jet-quenching parameter $\hat{q}$. For the absorptive part of the cross-section
$\Delta {\bf q}_{\perp}^{2}$ needs to be taken from data as in Eq.(\ref{avgperppiontot}) below or other theory.

We will take two reference energies for the jet, 20 GeV and 100 GeV. Both of them are high enough that two jets in $e^- e^+$ collisions would be clearly identifiable. But the interesting difference comes from the hadronization time. One can estimate a parton to have a lifetime in the laboratory frame, before hadronizing~\cite{Habel:1990tw} of order
\be \label{hadronizationtime}
\tau \sim \frac{1}{M_{\rm constituent}} \frac{E}{M_{\rm constituent}} \ ,
\ee
where the first factor is the typical scale of positivity violation in a colored-particle propagator and the second the time-dilation factor to the lab frame. For a gluon it is known from the lattice gluon spectrum and propagators as well as from Dyson-Schwinger studies that $M_{\rm constituent}\sim 0.8 $GeV, and $E\sim \sqrt{s}\sim P_t$.
Therefore, $\tau(100{\rm GeV}) \sim 30$fm while $\tau(20{\rm GeV}) \sim 6$fm. Therefore, the higher energy jet hadronizes well after the collision is over and behaves still as a parton if it enters the hadron gas, while the lower energy jet has a lifetime comparable to the size of the quark-gluon plasma and can enter the hadron gas already as a hadronized jet.

\subsection{Elastic and diffractive scattering}
First, let us consider the contribution from elastic scattering, so the number of particles does not change. Since the process is dominated by low momentum transfer, the largest contribution is the Pomeron's, and Regge analysis must be performed~\footnote{Estimates of the quark-pion cross section at low energies exist~\cite{Kalinovsky}, but are not too relevant for jet physics.}.
According to Regge theory, high energy elastic cross sections become
\begin{equation} 
\label{simplepomeron}
\frac{d \sigma_{\rm elastic}}{dt} = A \exp (b~t)
\end{equation}
with the $b$ exponent quite universal in hadron-hadron collisions. 
The slope is measured for energies of order $E=10$ GeV in proton-proton processes~\cite{Okorokov:2009em},
to be  $b^{pp}|_{10~{\rm GeV}} \simeq 10-11$ GeV$^{-2}$. It is about half for $\pi\pi$ 
interactions~\cite{Schuler:1993wr}, and we take
$b^{\pi\pi}|_{10~{\rm GeV}} \simeq 5-6$ GeV$^{-2}$.
Note the $s$-independence of the cross-section at the power-law level (we ignore the residual $\log^2(s)$ 
dependence typical of the pomeron since we will remain in the order of magnitude between 10 and 100 GeV).

We need to fix the normalization constant $A$ and for this we use 
the relation of Gribov-Pomeranchuk
\be\label{sigmapipi}
\sigma_{\pi\pi} = \frac{\sigma_{\pi p}^2}{\sigma_{pp}} \simeq 1.3 ~{\rm mbarn }
\ee
based on Regge factorization and  well-measured proton-proton and pion-proton cross-sections. Then, 
\be \label{totalcross}
A = b \sigma_{\pi\pi}  \ .
\ee
 
Using the differential cross section in Eq.~(\ref{simplepomeron}), the average squared momentum transfer is not very high
\begin{equation} \label{avgperppionel}
 \left<\Delta {\bf q}_{\perp}^{2}\right>_{\rm el} =  \frac{1}{b} \simeq (0.44~~\rm{GeV}) ^{+2}
\end{equation}
and, consistently, well within the soft Regge regime.

Another common phenomenon is the diffractive dissociation of either the target or the projectile or both, leaving a rapidity gap. The cross sections are now of order 3.4 mbarn (single diffractive on either particle) and 1.2 mbarn (double diffractive, breaking both target and projectile).
The $t$-dependence is analogous to Eq.~(\ref{simplepomeron}) in both cases,
\begin{equation} 
\label{simplepomerondif}
\frac{d \sigma_{\rm sd}}{dt} = A^{sd} \exp (b^{sd}~t)\ ; \ \ \ 
\frac{d \sigma_{\rm dd}}{dt} = A^{dd} \exp (b^{dd}~t)
\end{equation}
with $b^{sd}\simeq2.8$GeV$^{-2}$ and $b^{dd}\simeq 2$GeV$^{-2}$. For single diffraction the transverse momentum 
averages to
\begin{equation} \label{avgperppiondif}
 \left<\Delta {\bf q}_{\perp}^{2}\right>_{\rm diff} =  \frac{1}{b} \simeq (0.60~~\rm{GeV}) ^{+2}
\end{equation}
($(0.71{\rm GeV})^2$ for the double diffractive one).

\subsection{Total cross-section}

All other phenomena (the biggest part of the cross section at high energies) can be thought of as ``absorptive'' in the sense that the pion that interacted disappears from the outgoing beam, since the reaction is inelastic.

Nevertheless, the perpendicular momentum transfer is well defined as the ``opening'' of the jet cone; one tracks the longitudinal and transverse momentum and not the actual particle carrying it. 

But we need a different way of assessing the typical $\Delta {\bf q}_\perp$ per collision from data since the projectile disappears and cannot be used to define it. We resort to classic $\pi N$ scattering data~\cite{procCERN}. The data shows that the perpendicular momentum per emitted particle in an inelastic collision is quite independent of energy, multiplicity and nature of the particles colliding: it is rather a property of the strong interactions. In the reference energy range that we have chosen around 20 GeV, the average multiplicity is about 4.25 particles in the final state per binary collision, and the average transverse momentum is 0.33 GeV. Thus, a higher estimate of the momentum acquired by the 20 GeV pion, now turned into debris, is the product of both quantities, or about
\begin{equation} \label{avgperppiontot}
 \left<\Delta {\bf q}_{\perp}^{2}\right>_{\rm tot} =  \frac{1}{b} \simeq (1.4~~\rm{GeV}) ^{+2}
\end{equation}

\begin{figure}
\begin{center}
\includegraphics[width=0.50\textwidth]{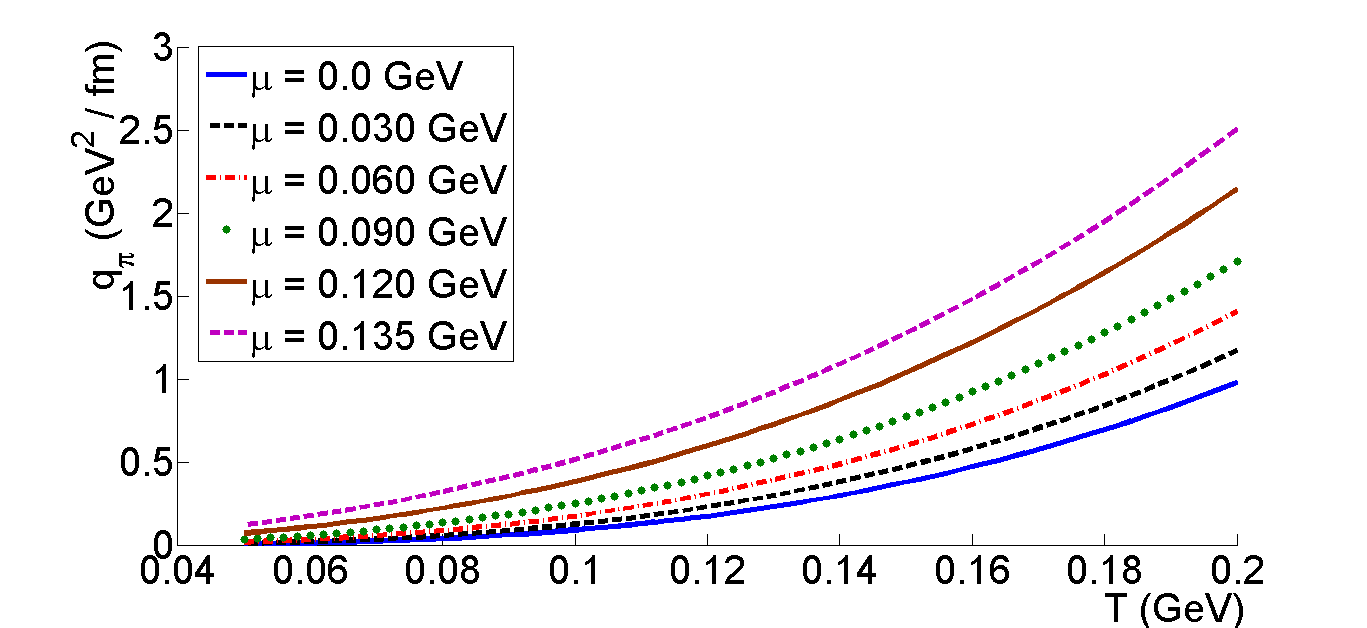}
\end{center}
\caption{Jet quenching parameter $\hat{q}=\hat{q}^{(1)}$ for one highly energetic meson entering a meson gas. We show the dependence with the gas temperature and (approximate) pion chemical potential $\mu$ (growing from bottom to top). There is no $P_t$-dependence because the pomeron-mediated cross section is largely independent of energy.}
\label{SolePion_vs_T}
\end{figure}

Considering only one pion in the jet, it therefore perceives a quenching parameter $\hat{q}^{(1)}$ as plotted in figure~\ref{SolePion_vs_T}. 

Since the pion gas cools upon expanding, we provide the temperature dependence (driven by the density as the cross-section is essentially energy-independent) and plot it for various chemical potentials that increase from bottom to top. Eq.~(\ref{BoseEinstein}) was  substituted, together with Eqs.~(\ref{sigmapipi}) and~(\ref{avgperppiontot}), into Eq.~(\ref{defqhat}) for the jet quenching parameter $\hat{q}$.

Since the dependence on $\mu$ plotted in figure~\ref{SolePion_vs_T} (and in the ones to follow) enters through the pion density alone, the jet quenching parameter is simply scaling in proportion to the fugacity $z=e^{\mu/T}$ that factors out of Eq.~(\ref{BoseEinstein}) with no other sensitivity to the chemical potential. This simple scaling is known from other systems~\cite{Dinmatter}  where the cross sections are relatively insensitive to the medium (moderate density). 
Unless we show the dependence on the chemical potential, we assume $\mu=0$ since available fits in the literature suggest that its value is small, of order 20 MeV~\cite{Torres-Rincon:2012sda,Stachel:2013zma}.

\section{Passage of a multiparticle jet through the hadron matter~\label{sec:hadrons2}}

\begin{figure}
\begin{center}
\includegraphics[width=0.50\textwidth]{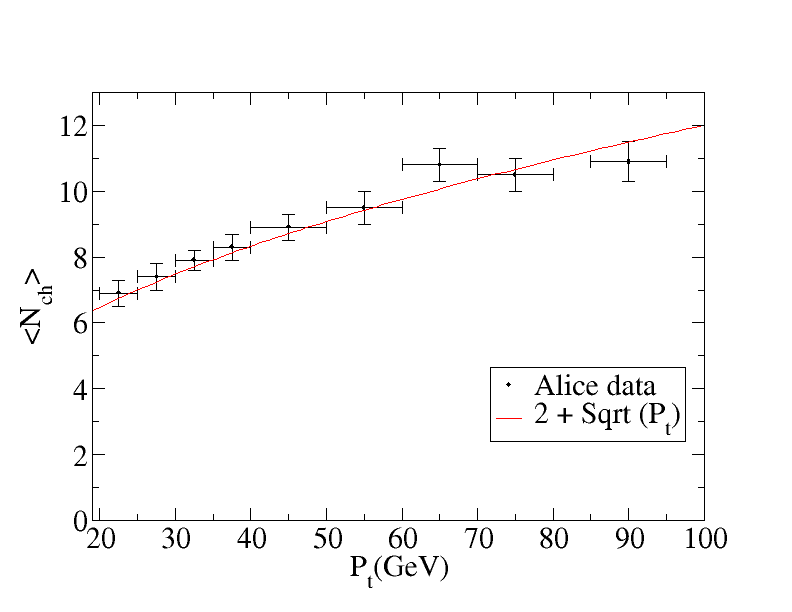}
\end{center}
\caption{Points with error bars: Alice~\cite{Prasad:2012vb} measurement of the average number of charged particles in a jet $\left<N_{ch} \right>$ as a function of the measured transverse momentum $P_{t}$. Solid line: fit by $\langle N_{ch} \rangle = 2+ \sqrt{P_t ({\rm GeV})}$.}
\label{NumParticulas}
\end{figure}

Having analyzed the transverse momentum acquired by a one-pion projectile, let us in the second place consider a jet of modest energy.
Due to the small time dilation, 
as discussed around Eq.~(\ref{hadronizationtime}),
the jet hadronized before entering the pion gas. 
Each jet's hadron interacts with the medium as in sec.~\ref{sec:hadrons}, 
but we need to multiply by the average number of particles in the jet to obtain $\hat{q}^{(N)}$.
We take this average number, among the various experiments that have measured charged jet multiplicity as a function of $P_t$, from the more recent ALICE collaboration data~\cite{Prasad:2012vb}, displayed in figure~\ref{NumParticulas}.
As can be seen, the data is well described by a fit of the form 
$
\langle N_{ch}(P_t) \rangle = 2+ \sqrt{P_t({\rm GeV})} \ . 
$
This is not unexpected, as multiplicities in $e^-e^+$ collisions typically also grow with the square root of the energy.

Neutral particles are less well characterized by experiment. Since a large fraction of the jet components are known to be pions (the lightest hadrons),  the ratio $N_0:N_{ch} = 1:2$ is natural,
so that a reasonable starting point is
\be \label{Ptdep}
\langle N(P_t) \rangle = \la N_0 + N_{ch} \ra = 3+ 1.5\sqrt{P_t({\rm GeV})} \ .
\ee
For the less abundant nucleons and kaons, the ratio $N_0:N_{ch}$ is rather $1:1$, so that the estimate typically undercounts the number of particles  at the 10\% level. 
Also the large-$E$, small-$P_t$ Kaon cross-sections are very similar to pion ones, so we need not distinguish them. The experimental cross sections~\cite{PDG} for nucleons are larger, empirically scaling with the number of constituent quarks, so by taking all jet and all medium particles as mesons we are underestimating the average cross section slightly. In all, our rough numbers should be taken with confidence at the 25\% level.

\begin{figure}
\begin{center}
\hspace{-1.3cm.}\includegraphics[width=0.53\textwidth]{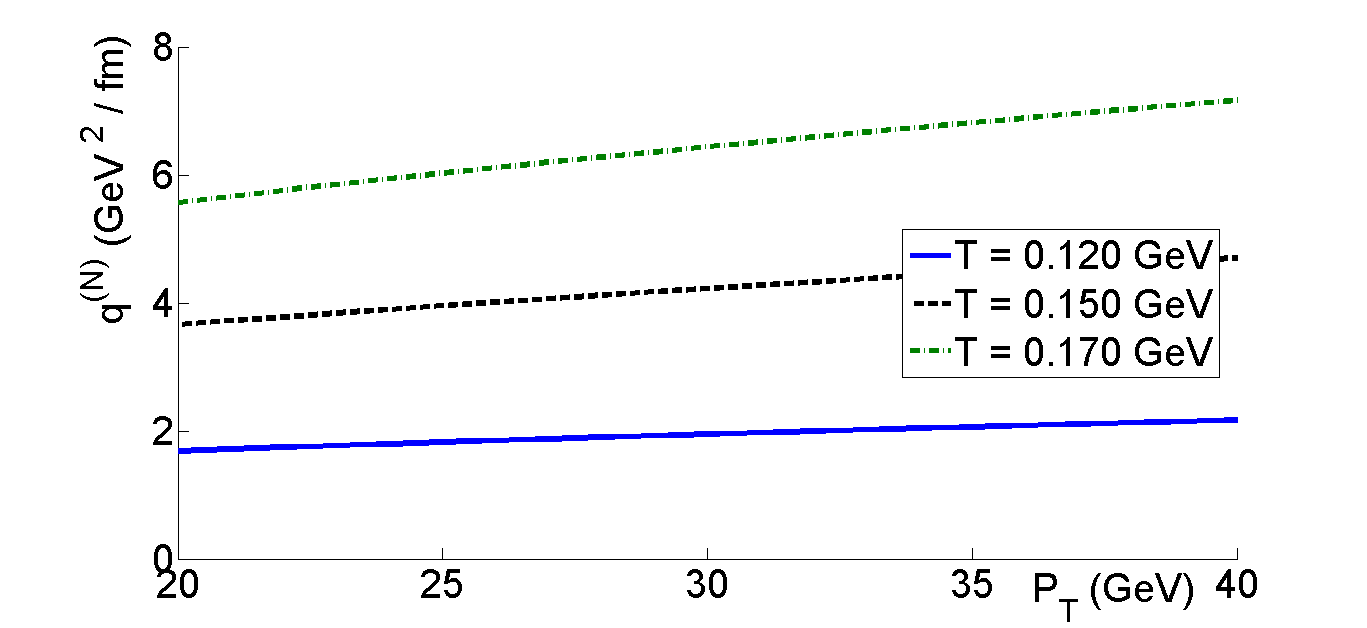}
\end{center}
\caption{Modest dependence of $\hat{q}^{(N)}$ in $P_{t}$ in the pion gas phase for moderate values of this variable. This $\sqrt{P_{t}}$-like dependence is caused by the softly increasing number of particles in the hadronized jet seen in Fig.~\ref{NumParticulas}. The temperature increases from bottom to top (120, 150 and 170 MeV respectively).
The chemical potential has been fixed at $\mu = 0.0$ GeV.
}
\label{Jet_vs_Pt}
\end{figure}

\begin{figure}
\begin{center}
\hspace{-1.0cm.}\includegraphics[width=0.50\textwidth]{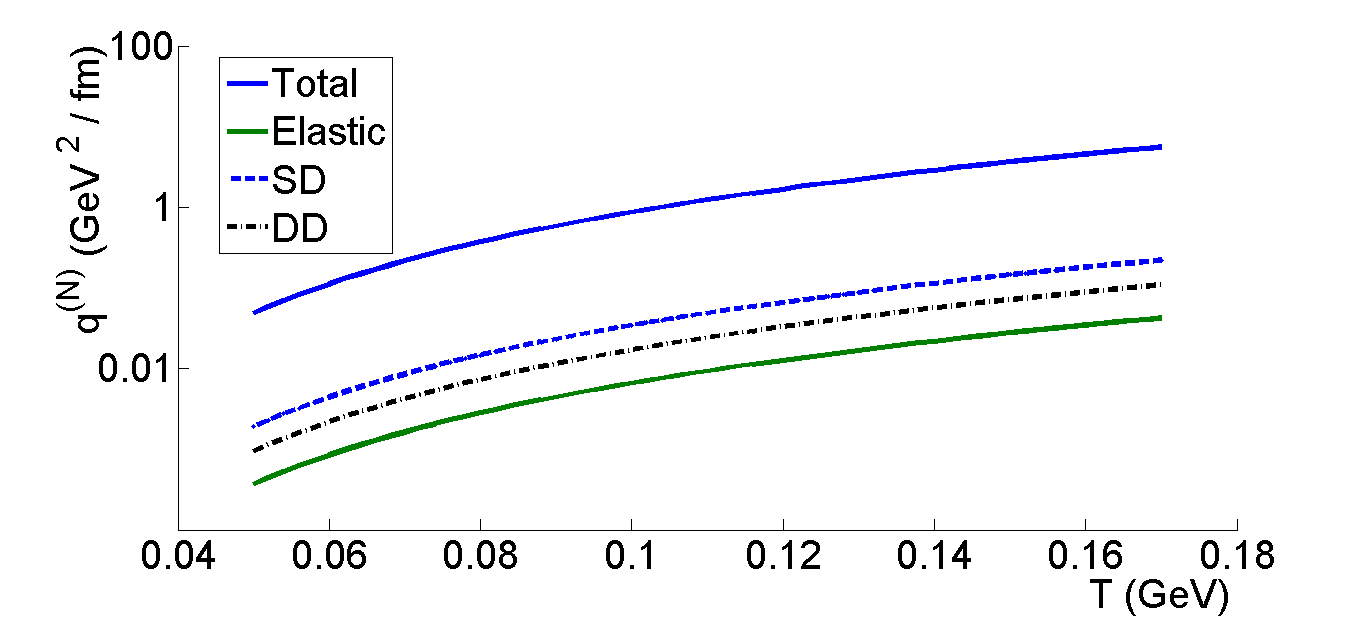}
\end{center}
\caption{Jet quenching parameter $\hat{q}=\hat{q}^{(N)}$ for a (hadronized) jet of energy around 20 GeV entering a meson gas as a function of the gas temperature.  From bottom to top, contributions of the elastic, double-diffractive, single-diffractive and total cross sections. We have fixed the pion chemical potential to $\mu = 0.0$ GeV. 
}
\label{fig:Total_plus_elastic_plus_diff_vs_T}
\end{figure}

Eq.~(\ref{Ptdep}) relating the number of particles in a jet to its transverse momentum induces a dependence of the jet quenching parameter in this variable,
$\hat{q}(P_t)$ (inasmuch as it is hard to distinguish $\hat{q}^{(1)}$ from  
$\hat{q}^{(N)}$ in experiment). Additional power-like dependence cannot be acquired from the total cross-section (because it is energy independent, $\sigma\propto s^0\sim \left( P_t^2\right) ^0$ ). Thus, quenching of low- to mid-energy jets in the hadron gas grows effectively with 
\ba \label{qsengas}
\hat{q}^{(1)} \propto P_t^0 \\ \nonumber
\hat{q}^{(N)} \propto \sqrt{P_t}\ .
\ea
 This last dependence is depicted in Fig.~\ref{Jet_vs_Pt}.  
Having established it, we can set for the rest of the analysis of lower momentum jets a fixed value $P_{t} = 20$ GeV.
In figure~\ref{fig:Total_plus_elastic_plus_diff_vs_T} we compare the contributions to the jet quenching parameter
coming from the elastic, diffractive and total cross-sections.
Because the total cross section (largely absorptive) is much larger than the elastic and the diffractive ones, and moreover the average ${\bf q}_\perp$ transfered is also larger, the resulting $\hat{q}$ is way bigger considering the total cross section. 
The number is large, of order 1-10 GeV$^2$/fm and thus of the same order or even larger
as experimentally measured values (though the pion gas, of course, is active only in the final stages of the collision).

Now let us address very energetic jets, of higher momentum of order $P_t=100$ GeV, that, due to large time dilation, have not hadronized when they exit the hot phase into the hadron gas and thus consist of one very energetic parton, entering the hadron gas and scattering off the pions there; we therefore need an estimate of the relevant cross-section to substitute into Eq.~(\ref{defqhat}).

The interaction between this colored parton and a pion in the gas cannot be obtained directly from experiment because of color confinement. We therefore need to resort to theory where the concept of a parton-hadron scattering amplitude~\cite{Landshoff:1970ff} can be exploited in spite of the partons not being in the asymptotic spectrum.

\begin{center}
\begin{figure}
\hspace{0.5cm}
\includegraphics[width=0.40\textwidth]{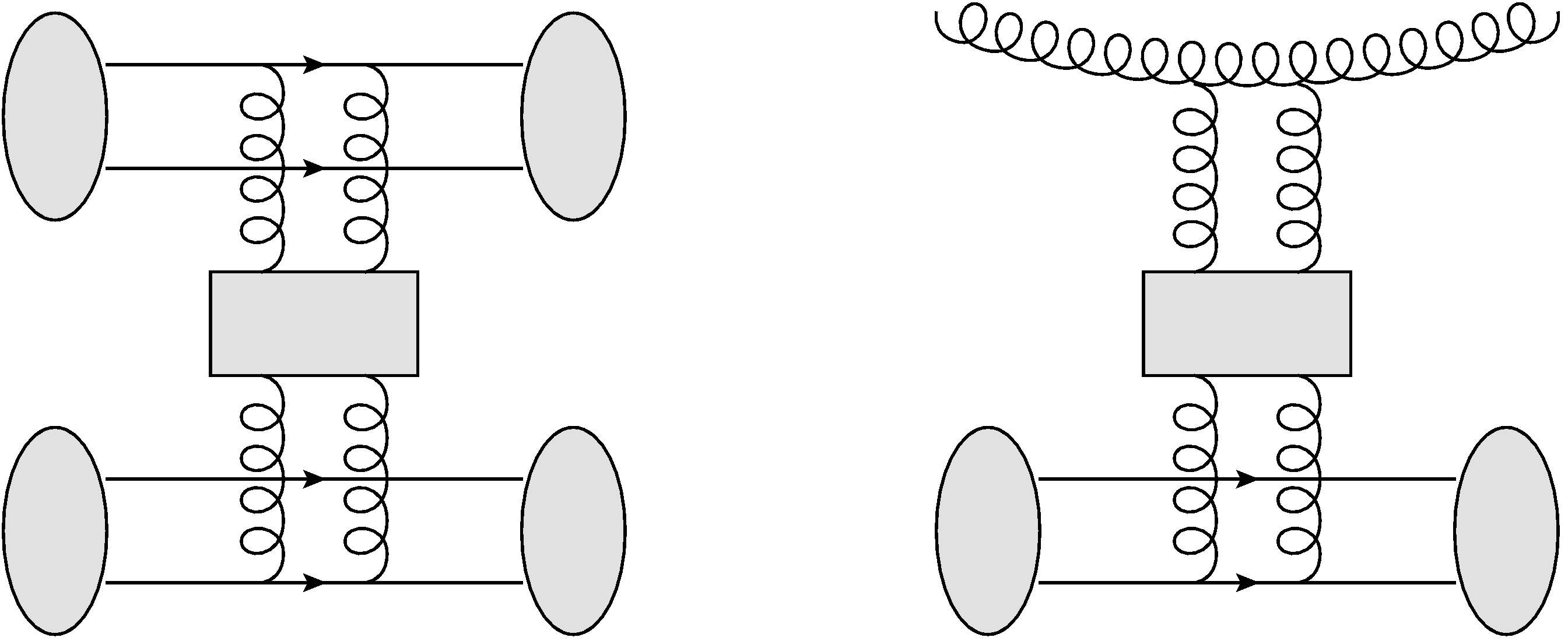}
\caption{Feynman diagrams showing pion-pion pomeron-mediated scattering (left) and similar gluon-pion scattering (right). They are of the same order in the large-$N_c$ counting.\label{colorfig}}
\end{figure}
\end{center}

In the first place we invoke the interpretation of the pomeron as an exchange of two gluons that rescatter multiple times (thus, the particles in the pomeron Regge trajectory are glueballs starting with spin and parity $2^+$~\cite{Bicudo}).
Then, the dependence on $s\simeq P_t^2$ is the same for both pion-pion and parton-pion scattering. The cross-section normalization however depends on the coupling of the two gluons that start the ladder to either source (see fig.~\ref{colorfig}). 

Second, we limit ourselves to the large-$N_c$ counting and note that the two diagrams are of the same order (differences would arise if we included baryons~\cite{HidalgoLlanes}).
Thus, in leading-$1/N_c$ accuracy (or 40\%), we cannot distinguish whether a parton or a meson entered the pion gas and we thus employ the same cross section for illustrational purposes.
Hence, a very energetic parton (that had no time to hadronize before entering the pion gas) acquires transverse momentum at a rate similar to that of a single pion. 

In conclusion of this section, a hard jet ($P_t$ in the hundreds of GeV) is kicked transversally in the pion gas as represented in figure~\ref{SolePion_vs_T}, corresponding to $\hat{q}^{(1)}$. A softer jet (with $P_t$ rather in the 10-20 GeV range)
instead, having hadronized and being composed of many particles, each interacting with the pions in the gas, acquires transverse momentum faster as represented in figure~\ref{Jet_vs_Pt}, that corresponds to $\hat{q}^{(N)}$.

\section{Jet quenching in the quark-gluon plasma/liquid~\label{sec:plasma}}

\subsection{Perturbation theory}

It is commonly accepted~\cite{Wang:2003aw}  that the observed jet quenching level is mostly due to parton scattering in the quark-gluon plasma. Though we find large values for $\hat{q}$ in the pion gas, this is short-lived at the end of the collision. Thus, and also for comparison, we briefly assess the coefficient in the color plasma phase.

Bjorken, who first treated the problem, did not initially appreciate the strong effect of  braking radiation (often off-shell, more analogous perhaps to pair production in QED) and assigned the loss of energy per unit length to the collisional processes. 
He thus estimated the energy loss of a parton 
with energy $E$ in a medium with energy density $\epsilon$ as
\be \label{bjorkensigma}
\frac{dE}{dx} \simeq CF \frac{\sqrt{30}}{2} \alpha_s^2 \epsilon^{1/2} 
\log \frac{2\langle k\rangle E}{M^2} \left( \frac{1+N_f/6}{1+21 N_f/32}\right)
\ee
based on the perturbative parton-parton scattering differential cross section at tree-level in QCD,
\be
\frac{d\sigma}{dt} = \frac{2\pi\alpha_s^2}{t^2}\times CF
\ee
with color factor $CF=2/3$ for quarks, $3/2$ for gluons. The flavor factor for $N_f=3$ is of order 1; The scale $M=1$ GeV has been later refined by other authors, for example~\cite{Braun:2006vd}, by properly handling interactions in the medium within field theory to interpret M as $\mu_s$, the Debye screening mass \footnote{We have substituted $M$ by $\mu_s(T)$ and found tiny differences in the computer code for $\hat{q}$, with at most 2\% change. This is of course because the scale $M$ enters only logarithmically, and the Debye mass for $T\in(300,500)$MeV is $\mu_s\in(0.6,1)$GeV, just a bit smaller than $M$.
}
; $\alpha_s=0.2$ the strong coupling constant (that we update here by running to the scale $P_t$ from the $Z$-pole value~\cite{PDG} at one-loop); 
and the average momentum $\la k\ra = 2T$.

\begin{figure}
\begin{center}
\hspace{-0.8cm}
\includegraphics[width=0.50\textwidth]{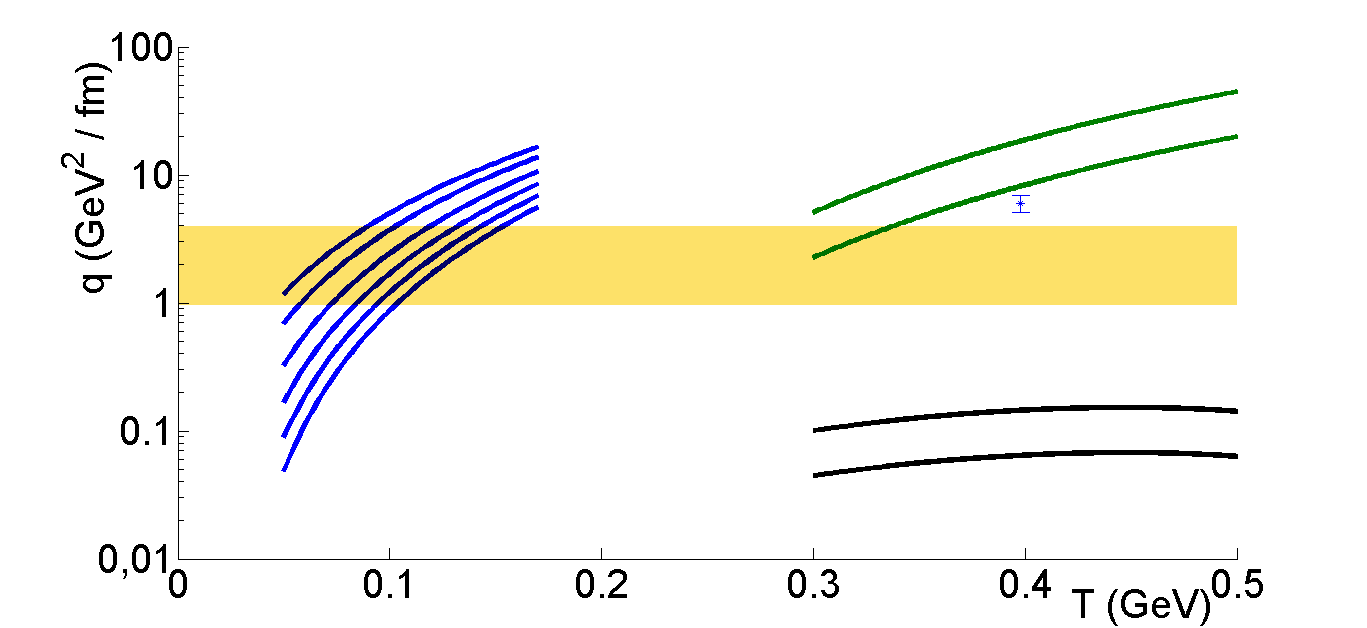}
\end{center}
\caption{$\hat{q}$ as function of temperature $T$ (at fixed transverse momentum $\left< P_{T} \right> = 20$ GeV). Low temperature lines to the left (blue online): hadron gas estimates for 
$\hat{q}^{(N)}$ for various chemical potentials, from bottom to top 
$\mu = 0, 0.030, 0.060, 0.090,0.120,0.135$ GeV. At higher temperature towards the right: Bjorken's (top lines, green online) and Arnold and Xiao's (bottom lines, black online) perturbative prediction for $\hat{q}^{(1)}$ from Eq.~(\ref{bjorkensigma}). For each pair, the top line corresponds to gluon- and the lower line to quark-energy loss. In addition, the lattice point from Panero, Rummukainen and Schäfer as well as the experimental extraction by the JET collaboration are shown.
}
\label{Low_Pt_vs_T}
\end{figure}

From today's perspective,  that collisional loss of energy in Eq.~(\ref{bjorkensigma}) is best quoted as an acquisition of transverse momentum via~\footnote{To obtain this equation we simply write $E^2=m^2+q^2_{//}+q^2_{\perp}$  and take the partial derivative of $E$ respect to $\ar {\bf q}_\perp\ar $. Since $q_{//}\simeq E$ does not vary significantly in the collision, the partial and total derivatives can be equated. We obtain 
$2E dE/dx = d(q_{\perp}^2)/dx$ from which Eq.~(\ref{qBjorken}) follows.
}
\be \label{qBjorken}
\hat{q}_1 \simeq 2E \frac{dE\ar_{\rm collisional}}{dx}
\ee
Then in figure~\ref{Low_Pt_vs_T} we replot Bjorken's computation in the higher temperature plasma phase against the hadron-gas estimate from section~\ref{sec:hadrons2} given in figure~\ref{fig:Total_plus_elastic_plus_diff_vs_T}.

\begin{figure}
\begin{center}
\hspace{-0.6cm}
\includegraphics[width=0.50\textwidth]{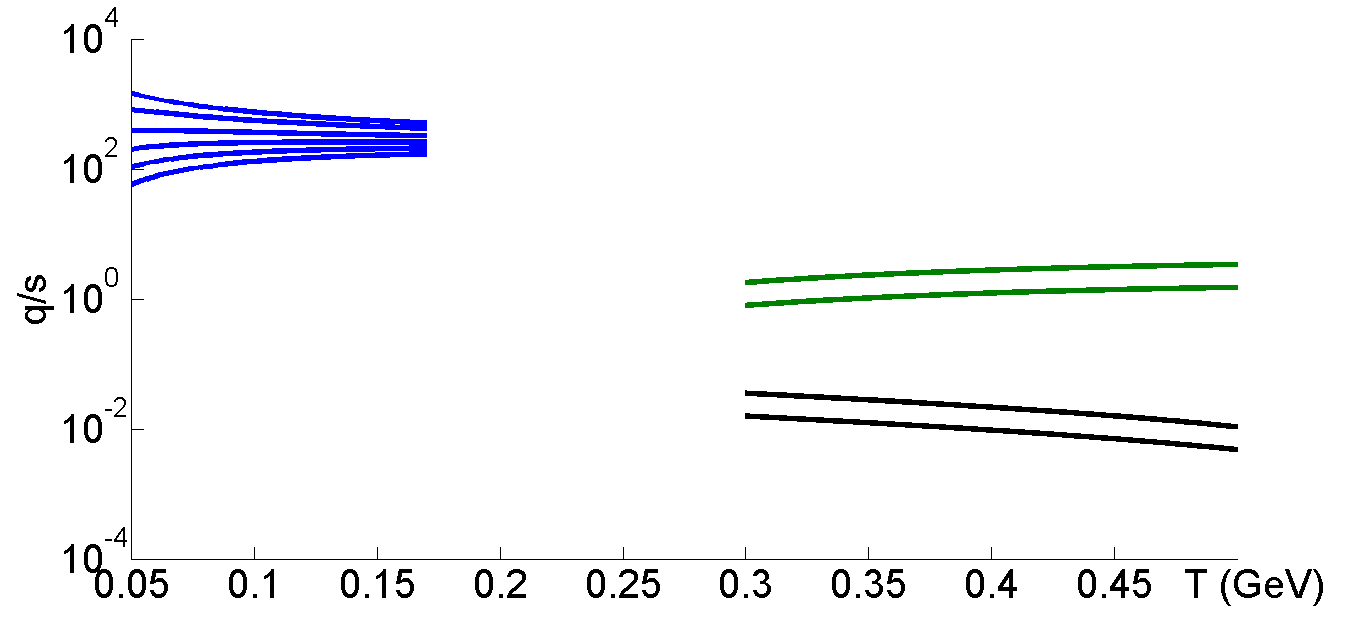}
\end{center}
\caption{
Same as figure~\ref{Low_Pt_vs_T} but normalizing the jet quenching parameter to the entropy density, $\hat{q}/s$, yielding an adimensional number. 
}
\label{qovers}
\end{figure}

Eq.~(\ref{defqhat}) is proportional to the density (as are the Bernouilli pressure and the drag force in classical fluid mechanics), so it is useful, when comparing disparate media, to normalize the transport coefficient as $\hat{q}/n$, if one is interested in learning about the underlying interactions from transport. However, since for viscosities the denominator is taken as the entropy density, we will do the same here (the ratio between number and entropy densities being a constant for an ideal gas).

Therefore, in figure~\ref{qovers} we plot the jet quenching parameter in adimensional form, that is, normalized by the entropy density also denoted as $s$ (do not confuse it with Mandelstam's $s$ as used in the rest of the article). We have used the ideal-gas formulae
\be
s= \frac{2\pi^2}{15} T^3
\ee
in the pion gas (with degeneracy factor $g=3$) and
\be
s= \frac{2\pi^2}{45} \left(16+ \frac{7}{8} 36 \right) T^3
\ee
on the quark-gluon plasma side (eight gluons with two polarizations, and three quark flavors with three colors and two spin components each). $\hat{q}/s$ is rather flat because the cross-section is quite independent of the energy. 
The left-side pion-gas lines (blue online) show higher values reflecting the difference between $\hat{q}^{(N)}$ and $\hat{q}^{(1)}$ (the jet is hadronized in the pion gas side, and $N\simeq 9$ particles lose energy faster than only one parton), and also that the perturbative cross-sections in the plasma are smaller than the pomeron cross-sections in the pion gas side.

We have also replotted in figures~\ref{Low_Pt_vs_T} and~\ref{qovers} a contemporary evaluation of the same phenomenon as analyzed by Arnold and Xiao~\cite{Arnold:2008vd} (see \cite{CaronHuot:2008ni} for higher order corrections). Their result for the jet quenching parameter is
\be \label{qArnold}
\hat{q} = CF \left( 6 ~I_{+} + 2\cdot N_{f}~I_{-} \right) \frac{g^{4}}{\pi^{2}} ~T^{3}
\ee
 with color factor $CF = \frac{4}{3}~ (3)$ for an incident gluon (quark), coupling $g = \sqrt{4\pi\alpha_{S}}$, and shorthand variables
\be
I_{\pm} = \frac{\chi_{\pm}(3)}{2\pi} \log\left(\frac{\Lambda}{M_{D}} \right) + \Delta I_{\pm}
\ee
with $\Lambda = 2$ GeV and $M_{D} = \sqrt{4\pi\alpha_{S}} ~T$ being two physical scales, 
$\chi_{+}(x) = \zeta(x)$ and $\chi_{-}(x) = \left( 1 - 2^{1 - 1/x}\right) \zeta(x)$. Finally, 
\ba
\Delta I_{\pm} = \nonumber \\
\frac{\chi_{\pm}(3) - \chi_{\pm}(2)}{2\pi} \left(\log\left( \frac{2T}{M_{D}}\right) + \frac{1}{2} - \gamma_{E}\right) - \frac{\Sigma_{\pm}}{2\pi}
\ea
in terms of the three constants $\gamma_{E}$ (the Euler-Mascheroni constant), $\Sigma_{+} \simeq 0.3860$ and $\Sigma_{-} \simeq 0.0112$, 
and of Euler's $\zeta$ function.

We do see quite some difference between the computation of Arnold and Xiao and that of Bjorken. The reason is that, barring logarithmic corrections, the newer result in Eq.(\ref{qArnold}) scales as $\hat{q}\propto T^3$, while Bjorken's $\frac{dE}{dx}\propto T^2$ in Eq.~(\ref{bjorkensigma}) leads to $\hat{q}\propto T^2 E$ in 
Eq.~(\ref{qBjorken}). Thus, the result of Arnold is smaller roughly by a factor $T/E$, which is small for a jet of 20 GeV as we consider here, and temperatures a few hundred MeV. 

We conclude from this subsection that if perturbation theory was all, $\hat{q}$ in the quark-gluon plasma would  be  smaller than in the hadron gas, and  insufficient for phenomenology.

\subsection{Lattice gauge theory}
Recently a lattice calculation of $\hat{q}$ at two temperatures, 398 MeV and 2 GeV has become available~\cite{Panero:2014sua}.
The computation proceeds by relating $\hat{q}$ to the integral of a collision kernel
$\int \frac{d^2 p_\perp}{(2\pi)^2 p_\perp^2 C(p_\perp)} $ that is the Fourier transform of a certain Wilson potential
$C(p_\perp) = \int d^2 x_\perp e^{ip_\perp x_\perp} V(x_\perp)$
that is calculable in the lattice by usual Euclidean methods in a reduced version of QCD, ``Electric'' or EQCD.
This effective theory is characterized by a high-temperature expansion that allows to keep only the zeroth Matsubara frequency and thus have one less dimension and an Euclidean metric. The drawback is that it is hard to make a statement about lower temperature physical systems. Some interesting consequences are $\hat{q}\propto T^3$ in that regime (as expected from dimensional analysis as commented above), an explicit calculation yielding
$\hat{q}(T=398{\rm MeV})\sim 6 $GeV$^2/$fm, and the observation that the lattice results are compatible with perturbation theory (in EQCD) if the perturbative expressions are taken at face value but substituting the Debye mass
(the infrared cutoff) from perturbation theory by a Debye mass computed in the lattice~\cite{Brandt:2014cka}.

The lattice point was included in figure~\ref{Low_Pt_vs_T}  above for perspective.

\color{black}

\subsection{Regge modeling}

\begin{figure}
\includegraphics[width=0.4\textwidth]{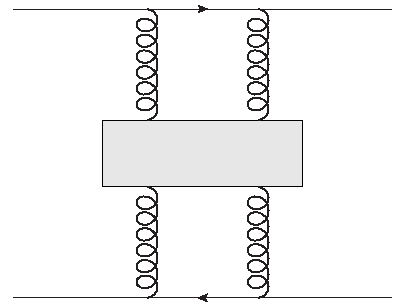}
\caption{\label{fig:pomeronplasma}
Parton-parton scattering at large $s$ but small $-t$ is a non-perturbative process presumably dominated by Reggeon exchange, independently of the parton virtualities $Q^2$.
}
\end{figure}

At temperatures of order few hundred $MeV$ perturbation theory is not to be trusted. Since momentum transfer is typically small but the jet energy is large, $s>>-t$, Regge kinematics applies.
We do not know of an actual Regge-like calculation from first-principles QCD, as $s$ is not so large that BFKL kinematics can be applied, so to provide estimates we have to avail ourselves of a soft-pomeron model (low $-t$) 
and the Mueller-Tang cross-section estimate based on resummation of elastic quark-quark scattering (large $-t$, but even larger $s$). The situation is depicted in figure~\ref{fig:pomeronplasma}.

\paragraph{Soft pomeron model}
A difficulty in making a soft-pomeron model is the lack of knowledge of the scale determining the $t$-slope $b$ in the QGP. In the example $\gamma^* p\to J/\psi +X$ the proton radius is known. In Regge theory for finite hadrons there is an empirical partition of the slope, 
\be
b(12\to 34) = b(1\to 3) + b(2\to 4) + b({\rm pomeron})\ ,
\ee
a consequence of Regge factorization, where
the first two terms reflect the structure and characteristic size of beam and target. 
For example, from photon-induced reactions we know~\cite{Ivanov} that, in GeV$^{-2}$, 
$b(\gamma p\to \rho p)=10\pm 1.5$,
$b(\gamma p\to \phi p)=7\pm 2$ and 
$b(\gamma p\to J/\psi p)=4.5\pm 1$, in the interval of interest for $t$ down to $-0.5$ GeV$^2$.
The logarithmic slopes $b$ for dissociative and elastic cross sections are not too different, with both cross sections differing more in the absolute normalization.

For our application of jet quenching, the ``beam'' is the hard parton, so $b(1\to 3)$ is very small, presumably of order $1/Q^2$ for a very virtual parton. The last, pomeron term, is universal~\cite{Donnachie}. The question is what to take for $b(2\to 4)$, that corresponds to whatever diffracting structures in the quark-gluon phase. One could conceivably try to evaluate the size and number of color-density fluctuations, perhaps through the fluctuation-dissipation theorem from knowledge of the color conductivity~\cite{Manuel:2004gk}. Or perhaps one should simply take the partons in the medium. 
In any case, we generally expect $b\ar_{\rm plasma}<b\ar_{\rm vacuo}$; but at high energy 
this decrease due to the lesser structure is partly compensated, for
we know that $b$ increases with $s$, as the hard pomeron slopes up in the Chew-Frautschi plot, 
\ba
s^{\alpha(t)} e^{bt} &\simeq & s^{\alpha(0) + \alpha' t} e^{bt} \nonumber \\
& = & s^{\alpha(0)} e^{(b+\alpha'\log s) t} 
\ea
(this phenomenon is the shrinkage of the forward cone).
For the purposes of this article we will ignore the issue and show results with $b_{\rm Soft~Pomeron} = 2.4$ GeV$^{-2}$ (see~\cite{yndurain}) to exemplify.  
The example differential cross section that we use, inspired by the soft pomeron, is 
\be \label{softpom}
\frac{d\sigma}{dt}\arrowvert_{SP} = \beta_{0}^{4}~ e^{bt} ~s^{\alpha t}
\ee
with $\beta_{0} = 1.32$ GeV$^{-4}$, $b = 2.4$ GeV$^{-2}$ and $\alpha = 0.2$;
(This purely phenomenological interaction can be modified for large virtuality to the hard pomeron's~\cite{Pirner2}).

\paragraph{Mueller-Tang resummation}
Another interesting Regge-based interaction occurs when $s\gg -t \gg \Lambda_{\rm QCD}^2$.
Since $-t$ is large, QCD reasoning is adequate, but because $s$ is much larger, the interaction is in the Regge regime.

We employ the results of Mueller and Tang~\cite{Mueller:1992pe}; 
in terms of $y\simeq \log\left( \frac{s}{-4t}\right)$, elastic quark-quark scattering proceeds with cross-section
\be \label{MuellerTang}
\frac{d\sigma}{dt} = \left( \alpha_s C_F\right)^4 \frac{\pi^3}{4t^2}
\frac{e^{(2\alpha_P-1)y}}{(\frac{7}{2}\alpha_s C_A \zeta(3) y)^3}
\ee
where as usual $C_F=4/3$, $C_A=3$, and $\alpha_s$ is evaluated at $-t$. The  pomeron intercept is $\alpha_P = 1+ 4C_A \log 2 \alpha_s/\pi$.

We compare these non-perturbative cross sections   with a typical perturbative one~\footnote{Reference~\cite{Kidonakis} is a good starting point for higher order formulae.
} for large $s$ and fixed $t$ with $\alpha_s=0.2$,
\be \label{pqcdsimple}
\frac{d\sigma}{dt}\arrowvert_{pQCD} = \frac{2\pi\alpha_{S}^{2}}{t^{2}}\ .
\ee
\begin{figure}
\begin{center}
\includegraphics[width=0.50\textwidth]{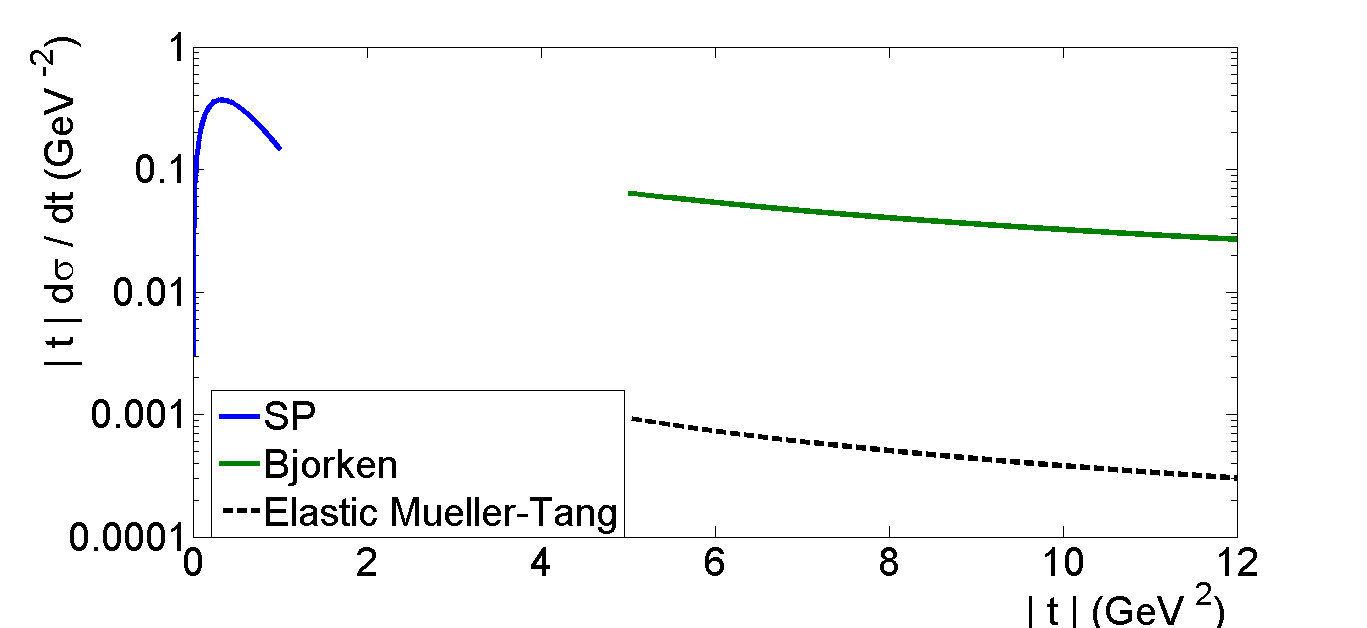}
\end{center}
\caption{Example elastic differential cross sections in the qgp phase as function of $|t|$, weighted by $|t|$. For small $|t|$ we show the soft-pomeron guess in Eq.~(\ref{softpom}) (blue online), and notice the usual exponential decay for small but increasing $|t|$. For large $|t|$, we show a tree-level perturbative QCD formula in Eq.~(\ref{pqcdsimple}) and the pqcd-Regge formula of Mueller and Tang~(\ref{MuellerTang}).
}
\label{High_Pt_vs_T}
\end{figure}
All three cross sections in equations (\ref{softpom}), (\ref{MuellerTang}) and (\ref{pqcdsimple}), weighted by $-t$, are shown  in figure~\ref{High_Pt_vs_T}. Indeed one induces peaking at low $t$, in total analogy with the $J/\psi$ production Hera data in figure \ref{Hera_Data_1}.

The elastic cross-sections (as the diffractive ones) are suggestive of low-t dominance, but it is the total cross section that will dominate mean free paths. Therefore we also quote the total Regge cross sections from resumming pQCD~\cite{barone}. These are, e.g. for quark-quark scattering
\be \label{totalqq}
\sigma_{\rm tot}^{qq} = \pi \frac{N_c^2-1}{N_c^2} \frac{\alpha_s^2}{{\bf k}_{\rm min}^2} \frac{e^{\lambda y}}{\sqrt{\pi \lambda'y/8}}
\ee
and for gluon-gluon scattering
\be
\sigma_{\rm tot}^{gg} = 4\pi \frac{N_c^2}{N_c^2-1} \frac{\alpha_s^2}{{\bf k}_{\rm min}^2} \frac{e^{\lambda y}}{\sqrt{\pi \lambda'y/8}}\ .
\ee
In these expressions $\lambda= (4\log 2) N_c \alpha_s/\pi$, $\lambda'=7\zeta(3) N_c\alpha_s /\pi$, $y=\log\frac{s}{{\bf k}_{\rm min}^2}$ and ${\bf k}_{\rm min}^2$ is the minimum transverse of the jet, a soft scale that for our estimates will be taken as $M^2 \sim 1{\rm GeV}^2$ also. 
Though in principle this infrared cutoff should also depend on temperature as was the case for the Debye screening mass in Bjorken's formula, the dependence is only logarithmic (and from our study above, small) so we drop it altogether.

\begin{figure}
\begin{center}
\includegraphics[width=0.50\textwidth]{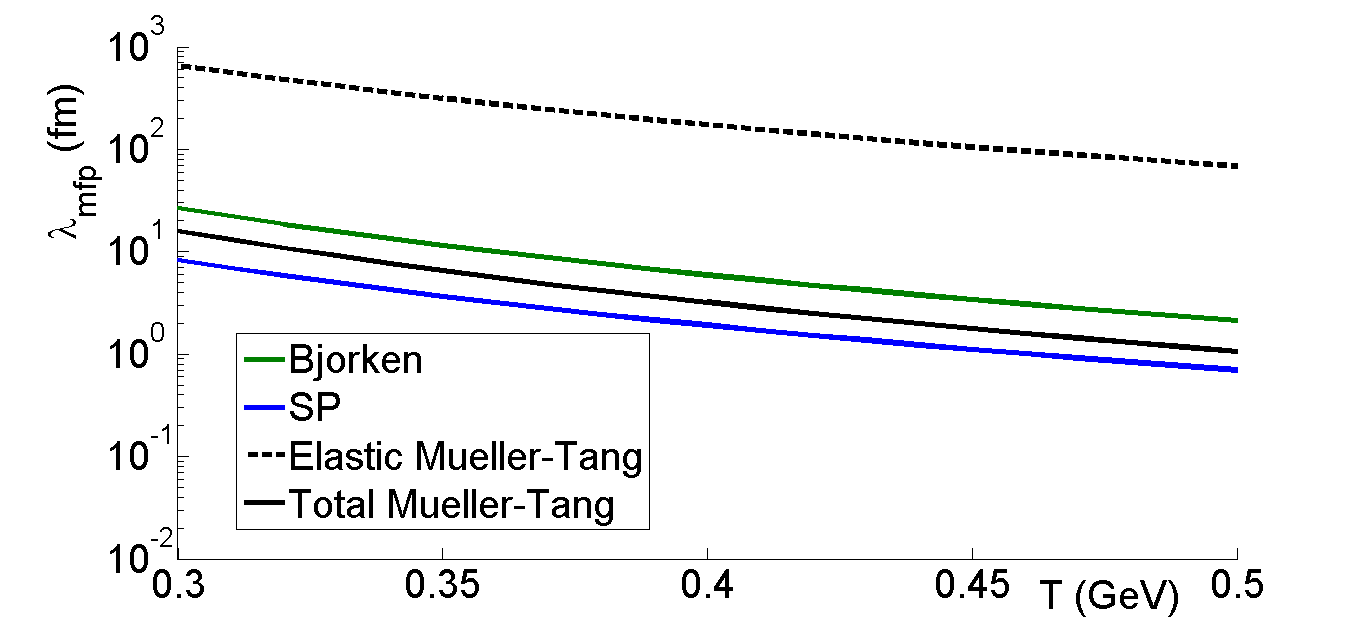}
\end{center}
\caption{Mean free path as a function of the temperature associated with each of the cross sections in figure~\ref{High_Pt_vs_T}, plus the total 
cross-section in equation~\ref{totalqq}. 
The mean free path associated with the total cross-section in pomeron models (two bottom lines, soft-pomeron and 
Mueller-Tang resummation) is quite smaller than for the perturbative interaction (third line from bottom).}
\label{High_Pt_vs_T_path}
\end{figure}

The corresponding mean free path for a high $P_T$ parton of 100 GeV is pictured in figure~\ref{High_Pt_vs_T_path} together with the ones infered from the total cross-sections integrated from equations (\ref{softpom}), (\ref{MuellerTang}) and (\ref{pqcdsimple}).
The figure shows that for high temperature, every couple of fermi through the plasma, a parton will be kicked transversally due to the soft interactions, while the perturbative (and elastic Mueller-Tang) ones have a mean free path at moderate temperatures that is not strong enough for all partons to be scattered transversally during the finite lifetime of the fireball.

\begin{figure}
\begin{center}
\includegraphics[width=0.50\textwidth]{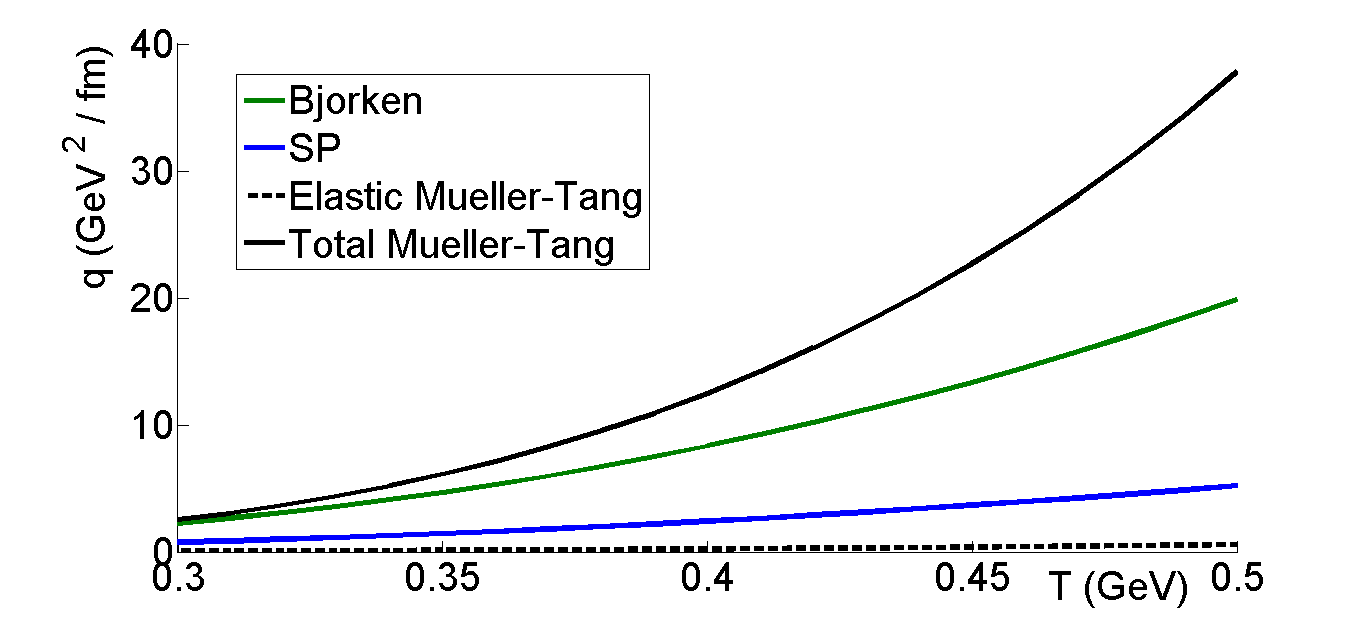}
\end{center}
\caption{From top to bottom, contributions to $\hat{q}$ attributable to total Mueller-Tang cross-section, elastic perturbative interaction, elastic soft pomeron and elastic Mueller-Tang in the QGP region, for $P_{t} = 20$ GeV, and following Eqs.~(\ref{softpom}, \ref{MuellerTang}, \ref{pqcdsimple}). 
A pomeron-mediated contribution would appear to be, at least, as important if not more than the perturbative QCD one, that has been extended to low momentum where it should be replaced by a full non-perturbative calculation, so it is an overestimate of the perturbative contribution.}
\label{High_Pt_vs_T_q}
\end{figure}

In figure~\ref{High_Pt_vs_T_q} we then show the corresponding contributions to $\hat{q}$ that would come about if the {\emph{entire}}
$t$ range of momentum transfers was effectively dominated by the corresponding process (so that the perturbative estimate is actually an upper bound of what perturbative interactions can actually contribute). It is clear that Regge interactions (note the top curve corresponding to Eq.~(\ref{totalqq})) can notably enhance the perturbative calculation of $\hat{q}$ and future work on the issue is warranted.

\section{${\bf q}_\perp$ correlations within the jet}\label{sec:correlations}
We have argued with some certainty that soft scattering dominates the jet quenching parameter in the hadron phase, but our knowledge of the quark-gluon phase is very primitive since it has turned out not to be a weakly coupled plasma. 

If, indeed, Regge theory plays also a role in the quark-gluon phase as 
we consider in this article, there should be some distinct phenomenological consequences.
First, we can discuss the scattering-energy (that is, jet-$P_t$) dependence of $\hat{q}$. 

The total cross section in Eq.~(\ref{totalqq})
gives for individual collisions, with $s\simeq P_t^2$,
$\sigma\propto \left(P_t^2\right)^{4(\log 2) N_c \alpha_s/\pi} \simeq \left(P_t^2\right)^{0.5}$ (for $\alpha_s$ around 0.2).

Comparing with the weaker dependence on the pion gas side, Eq.(\ref{qsengas}), we see that 
\ba \label{quenchpt}
\frac{d \log\hat{q}_{\rm qgp}}{d\log P_t}    & \simeq & 1 \\ \nonumber
\frac{d\log\hat{q}_{\rm hadron}}{d\log P_t} &    \in & (0,0.5) 
\ea
implying that more energetic jets lose energy faster, and especially so in the quark-gluon plasma side.

This is interesting since we already know from experiment that the suppression in the number of jets appears to be rather flat as function of $P_t$,  implying that more energetic jets have to lose more energy to be damped in equal numbers as the less energetic ones. Thus 
the jet-quenching parameter has to behave as in Eq.~(\ref{quenchpt}), with the larger exponent loosely favored by experiment.

\begin{center}
\begin{figure}
\includegraphics[width=0.4\textwidth]{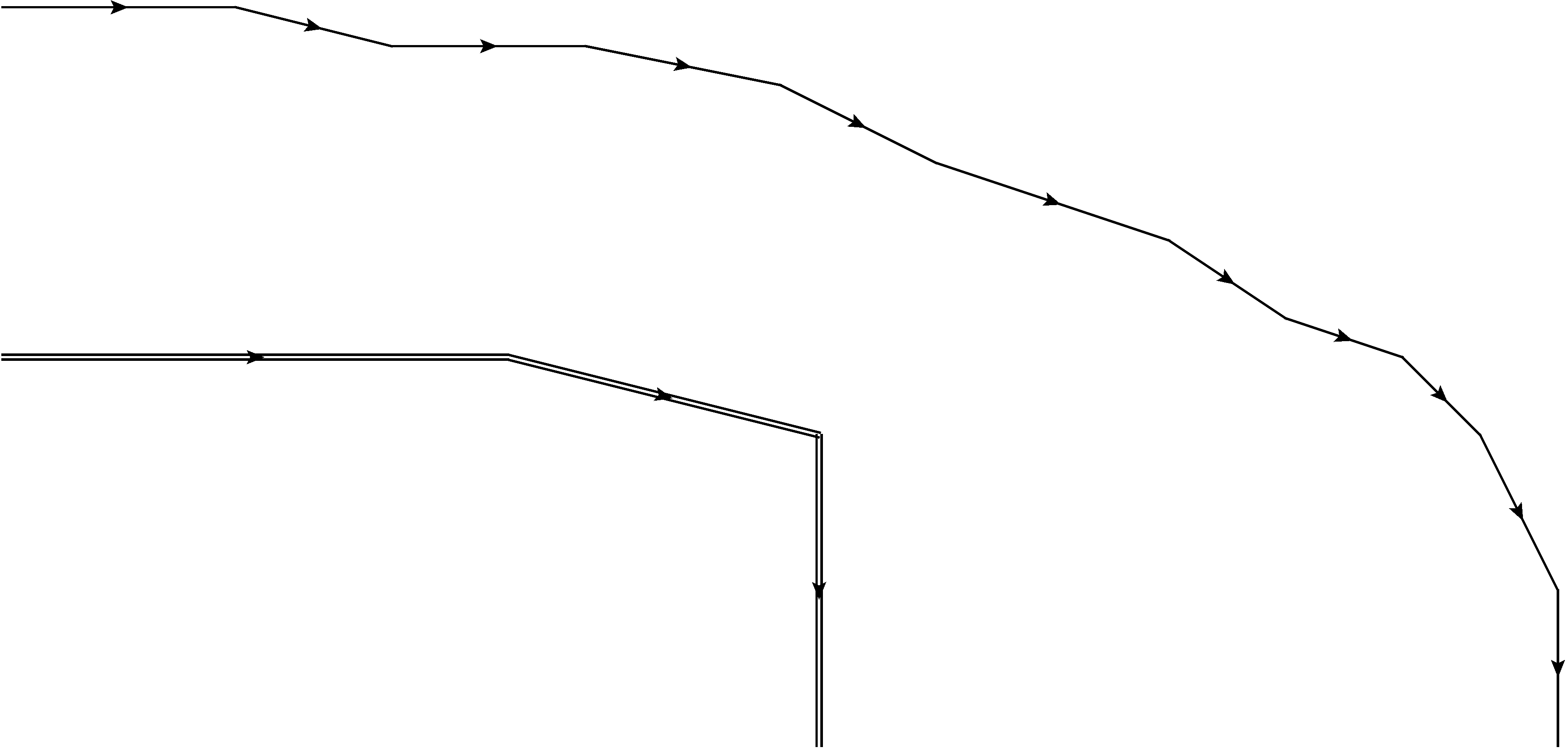}
\caption{A parton can acquire transverse momentum by multiple soft collisions with the medium (top, thin line), as in Brownian motion, or through very few harder collisions (bottom, double line).
A L\'evy flight would follow the top line with unfrequent longer jumps. \label{fig:HardorSoft}}
\end{figure}
\end{center}

We would like to propose additional experimental tests to distinguish between
the two mechanisms to achieve a given experimental ${\bf q}_\perp$ of the jet, by a few hard or by many soft collisions (see figure~\ref{fig:HardorSoft}).
What we propose is to measure the momenta ${\bf q}_{\perp i}$ of the hadrons in each jet and combine them to produce simple correlators that illuminate the size of the momentum transfer (the observables are not tailored to ascertain whether the collisions took place in the quark-gluon plasma or in the hadron phase, but rather the distribution of their intensity). With the ${\bf q}_{\perp i}$ we construct two simple event by event quantities whose distribution should be plotted (here, we simulate it).

\begin{center}
\begin{figure}
\includegraphics[width=0.50\textwidth]{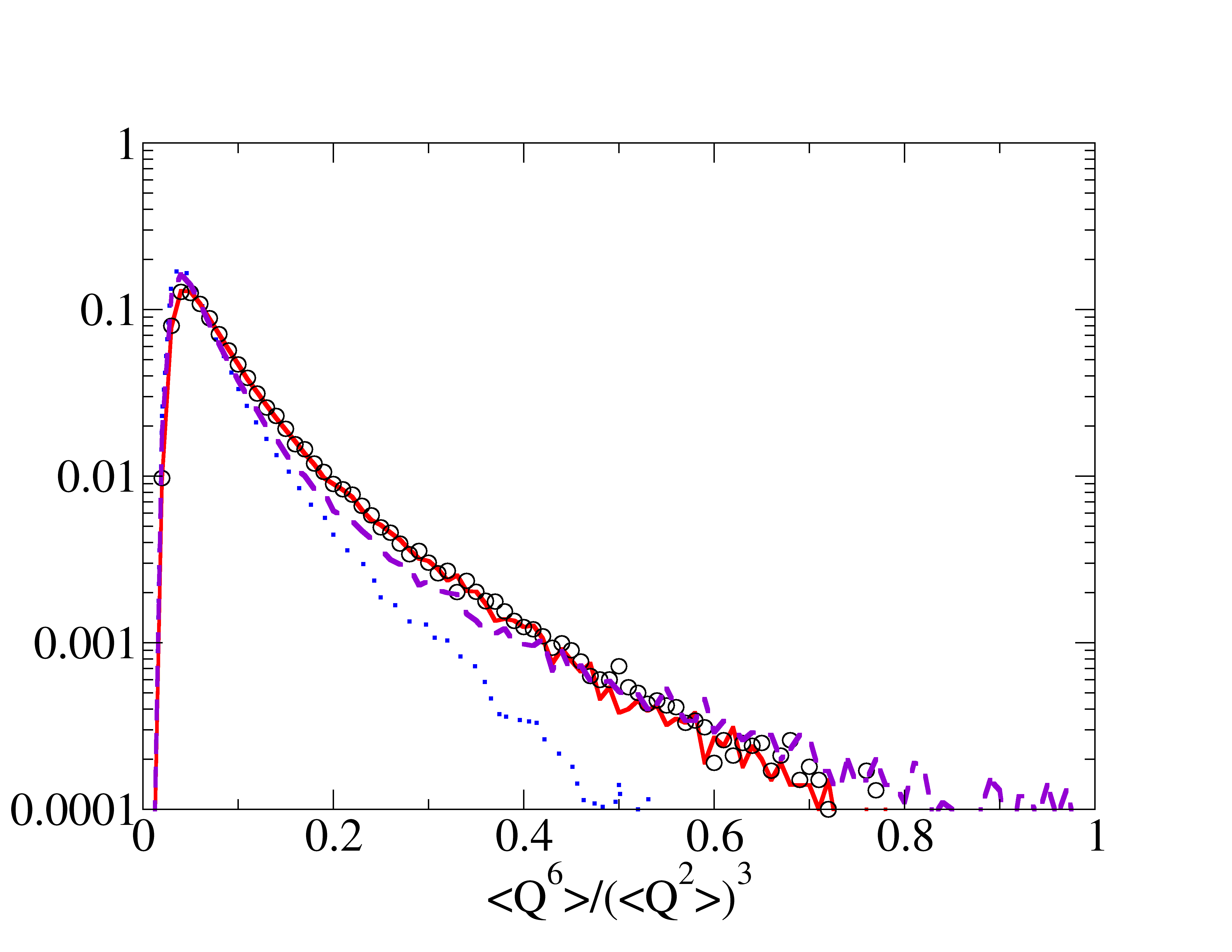}
\caption{We plot a (100-bin, normalized to 1)
histogram of $C_1=\sum_i {\bf q}_{i\perp}^6 /(\sum_i {\bf q}_{i\perp}^2)^3$ for various distributions of $d\sigma/d{\bf q}_\perp^2$ (see text). The right-most one (dashed line, violet online) has each momentum transfer distributed by a perturbative QCD tail, while the left-most dotted curve (blue online) comes from a pomeron-like soft function.
\label{fig:Q6}}
\end{figure}
\end{center}

We focus on 1 and 2-particle ${\bf q}_\perp$ correlations within the jet. 
In order to distinguish between different distributions of the interaction hardness in the medium, one needs to put different statistical weight on the hard ones than in the softer ones. Mathematically, power-law functions 
$f(x)=x^n$ with $n>0$ are way larger for $x\simeq 1$ than $x\simeq 0$. 
So it is natural to choose a dimensionless ratio with a high power (for example, $n=6$) 
\be
C_1 = \frac{ \sum_i {\bf q}_{i\perp}^6}{(\sum_i {\bf q}_{i\perp}^2)^3}
\ee
(the sum running over all hadrons in a jet) that enhances the (possible) few hard collisions. We plot it in figure~\ref{fig:Q6}.

To prepare the figure we have written a simple Monte Carlo simulation of 
$10^5$ swarms with exactly 10 particles each, initially collinear (all ${\bf q}_{\perp i}={\bf 0}$) and then kicked each particle at random in the transverse direction until a predetermined $<P_t^2>$ is reached, with $\Delta {\bf q}_{\perp}^2=-\Delta t$ distributed, from left to right in the figure (softer to harder), in proportion to each of the following four functions (units in GeV),
\ba 
& &8.8 e^{2.4 t}\ ; \nonumber \\
& &8.8 e^t \ ;  \nonumber \\
& &\left\{ \begin{tabular}{cc}
        $8.8 e^t$            &  low  t \\   
        $2\pi\alpha_s^2/t^2$ &  high t \\
\end{tabular} \right\}; \nonumber \\
& &2\pi\alpha_s^2/t^2\ .  \nonumber
\ea
(the factor 8.8 comes from demanding continuity of the third one).

\begin{center}
\begin{figure}
\includegraphics[width=0.50\textwidth]{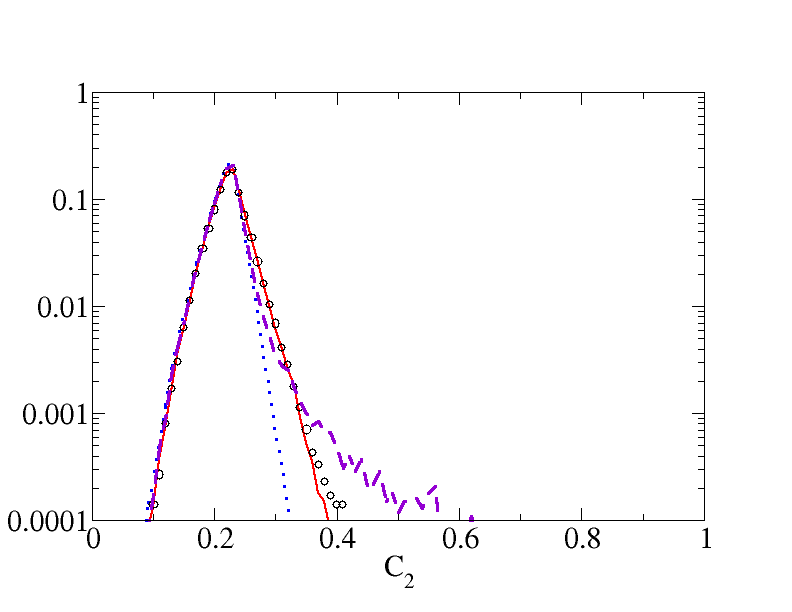}
\caption{(Lines as in figure
\ref{fig:Q6}.) We plot a (100-bin, normalized to 1) histogram with the distribution of
the correlator $C_2=\sum_{i\not = j} ({\bf q}_{\perp i}-{\bf q}_{\perp j})^2 / \sum_i{\bf q}_{\perp i}^2$. 
\label{fig:correlator}}
\end{figure}
\end{center}

Likewise, in figure~\ref{fig:correlator} we have done the same for a two-particle correlation with a simple geometric interpretation in the ${\bf q}_\perp$ plane (the sum of the distances between all pairs of particles divided by the sum of their distances to the origin), namely
\be
C_2= \frac{\sum_{i\not = j}({\bf q}_{\perp i}-{\bf q}_{\perp j})^2}{ \sum_i{\bf q}_{\perp i}^2} \ .
\ee

In both cases, if the individual particle collisions are distributed as a purely perturbative cross-section, $d\sigma/dt = \frac{2\pi\alpha_s^2}{t^2}$, the correlators can take high values with a characteristic ``fat tail''.
If the interactions were only soft one would have Brownian motion, with both correlators peaked at low momentum. 

In the more physical case with a soft interaction but also an ocasional hard collision, the random walk in the ${\bf q}_{\perp i}^2$ variable presents characteristics of a L\'evy flight~\cite{Bouchaud}. 

Indeed at fixed total energy $P_t$, the average momentum transfer to the particle   
\be
-\la {\bf q}_\perp^2 \ra =  \la t \ra = \int t \frac{d\sigma}{dt}
\ee
is logarithmically divergent due to occasional hard collisions. 
But since the momentum transfer $t$ is bound by $-s\simeq -P_t^2$, in rigor the distribution causing the L\'evy flight is truncated due to the finite energy in any experiment, and the flight is only a first approximation (although probably better than Brownian motion, with a Gaussian probability distribution in ${\bf q}_\perp$, as proposed in~\cite{D'Eramo:2010ak}).
Another reason to expect the L\'evy flight to only be a first approximation in a heavy ion collision is the fact that we are not dealing with infinite matter, but with a finite medium~\cite{Armesto:2011ht}.

We expect the correlators in a physical experiment to have some events in the mid-region but not quite near 1 since the physical interactions are only seldom hard.

Finally, if the interactions of the fast particles were purely soft, the correlators would have no tail at all (essentially no particle wanders away from the swarm due to hard collisions being extremely unlikely).

Thus, experimentally constructed final state momentum distributions within a jet can give information about the interactions of the fast partons with the medium. We believe that further theoretical investigations, particularly full Monte Carlo simulations with parton shower and simulated hadronization, should be the next step.

\section{Discussion}\label{sec:conc}
The body of literature on Regge theory, and also on jet quenching, is now inmense; this contribution not being a review, we have  highlighted only a few key and related works in the QGP to contrast with our own pion gas results.
Suffice to state again that in the simplest treatment, the first effect  triggering all subsequent phenomenology is controlled by the so called
\emph{jet quenching parameter} $\hat{q} = \Delta {\bf q}_\perp^2/\lambda$.

This parameter is proportional to the perpendicular momentum transfer squared, but in spite of this it is dominated by soft processes because they so much control the cross section.

We think that we can state with certainty that this is correct for a jet transversing the hadron gas, with the only caveat being the number of actual particles in the jet: this depends on the hadronization time (hence, $P_t$) and shower properties. We have simply used the experimental average number of particles to define an effective quenching parameter $\hat{q}^{(N)}$ that appears as $\hat{q}$ to the external observer.

An estimate of the jet quenching parameter in the pion gas exists~\cite{Chen:2010te}, but is way smaller (quoting $\hat{q}~(T=100~{\rm MeV})\simeq 0.5~ T^3$ instead of our $10-100~ T^3$). The reason must of course be the use of very different cross sections: while we have resorted to larger pomeron-like interactions, the (in principle) more sophisticated earlier work is based on purely perturbative ones~\cite{DengWang} through DGLAP splitting functions in the nuclear medium scaled to the hadron gas. We believe that in spite of the naivety of our estimate, it must be closer to reality than the smaller, existing appraisal.

Normally one would not be able to distinguish the effect of the pion gas under the presumably larger quenching in the quark-gluon medium and around the phase transition, but the RHIC beam energy scan, by proceeding to lower energies, may offer an opportunity by putting more weight on the later stages of the collision, and we look forward to data  analysis. 

Turning to the hotter quark phase,
the community is intensely dedicated to improving the perturbative treatment, and for example a recent lattice calculation~\cite{Panero:2014sua} reports   $\hat{q}_{\rm soft} \simeq 6$ GeV$^2/$fm for temperatures $T=398$ MeV, which is in the correct ballpark for phenomenology. 

Other recent works have pointed out 
that one can get higher values within perturbation theory~\cite{Zapp:2012ak},
or invoke more exotic mechanisms such as jet collimation~\cite{CasalderreySolana:2010eh} or
transfer of energy via plasma instabilities~\cite{Mannarelli:2009pd}.

Other authors~\cite{Wang:2003aw} are of course aware that, if a fast pion enters the pion gas, it will scatter softly. What we point out is that the same phenomenon is expected if an unhadronized parton enters the pion gas, and moreover, it is likely that the parton-parton interactions in the hotter phase of the quark-gluon plasma must also be of Regge origin. This makes the more urgent to understand how the pomeron behaves in a medium: though there are initial works on the pomeron at finite temperature based on QCD's BFKL equation~\cite{deVega}, and some old lattice references~\cite{Henty:1995qr,Parrinello:1996sk}, they seem to still be too far from being of practical use. There now is a gauge-independent way of obtaining the jet quenching parameter $\hat{q}$ 
(see~\cite{Liang:2008vz}, and recently~\cite{Benzke} for the derivation within the effective theory) that could perhaps be tried on the lattice, 
because it expresses $\hat{q}$ in terms of Wilson lines, as suggested in \cite{Cherednikov:2013pba}  (and this might also be a future way forward for Regge studies as there are expressions for pomeron exchanges in terms of these objects that have been sketched in the literature~\cite{Donnachie}).

Perturbative expressions such as Eq.~(\ref{pqcdsimple}) are generalized now by incorporating log resummations in the BFKL approach~\cite{DelDuca}.
An interesting development of Soft Collinear Effective Theory, called SCET-G~\cite{Idilbi:2008vm,Ovanesyan:2011xy,Ovanesyan:2011kn,Bauer:2010cc}, has included the effect of Glauber gluons (soft, Coulomb-like gluons) getting a step closer to the correct kinematics for Regge phenomenology. 
Nevertheless color-singlet pomeron exchanges seem absent from the analysis. 
Further work seems necessary to us and to others~\cite{Donoghue}.

In a beautiful application to jet quenching, D'Eramo, Liu and Rajagopal~\cite{D'Eramo:2010ak} have established a random walk in the ${\bf q}_{\perp}$ plane, though they include only soft collisions and thus obtain Brownian motion (with a characteristic Gaussian distribution) instead of a L\'evy flight with anomalous diffusion (possibly closer to QCD because, though infrequent, there are hard collisions).

The large number of observables studied nowadays in the context of jet quenching~\cite{Spousta:2013aaa} gives hope that perhaps one can be found that can distinguish between perturbative, Regge-like, and other mechanisms underlying momentum transfer to partons in the medium directly from experiment. We have proposed two examples.
From this viewpoint, we conceive that SCET or lattice-based efforts aimed at understanding this transport parameter will become more phenomenologically successful when they incorporate the Regge regime.

We have one more important piece of phenomenology to discuss.
From a geometric scaling analysis, Liao and Shuryak concluded that 
the jet-quenching parameter $\hat{q}$ probably has a maximum at or around the phase transition~\cite{Liao:2008dk}. This is not surprising since, in a gas, it is inversely proportional to the shear viscosity, that is known to have a minimum around the same phase transition. Additional support is given for example by~\cite{Domdey}, invoking resonance absorption near the critical temperature.

From the side of the hadron gas phase this possibility is not disfavored by our results: 
we find a non-negligible $\hat{q}$ in the hadron phase that grows distinctly with temperature towards $T_c$
and it might well reach a maximum. But it is hard to bracket this from the quark-gluon plasma side, even if we attend to the relatively small perturbation-theory calculations in figures 
\ref{Low_Pt_vs_T}, \ref{High_Pt_vs_T}, etc. And uncertainties in the plasma phase concerning soft scattering are also just too large to make definite statements. But if we were to dare a suggestion, it would appear that 
at LHC energies the effect of the maximum would be less visible in the data because of the hotter and longer plasma phase~\cite{Liao2}, and likewise
very high-$P_T$ jets (in the range of hundreds of GeV) would not present this maximum, while lower-$P_T$ (in the tens of GeV) effectively would, since they should be sensitive to $\hat{q}^{(N)}$ rather than $\hat{q}^{(1)}$ due to hadronization. Further work is also warranted here.

\section*{Acknowledgments}
We thank J. Liao from Indiana University for pointing to us that the literature lacks assessments of  $\hat{q}$ in the hadronic phase, which this work attempts to supply, and for reading the first manuscript, as well as M. Benzke, N. Brambilla, A. Gomez Nicola, A. Sabio-Vera and J. Torres-Rincon  for useful comments and references.
Financial support by Spanish Grants FPA2011-27853-C02-01 and FIS2011-28853-
C02-02. C. H.-D. thanks the support of the JAE-CSIC Program.


\end{document}